# The Profiling Potential of Computer Vision and the Challenge of Computational Empiricism

Jake Goldenfein[*]

'Profiling' means different things in different traditions. The proliferation of machine learning and data science in profiling however, make the European General Data Protection Regulation (**GDPR**) definition especially meaningful.[1] Under that regime, 'profiling' includes the automated evaluation of certain traits about a person,[2] reminding us of the legal significance of 'classification' even without subsequent discrimination. This definition also applies to an emerging class of automated image-based profiling technologies that evaluate 'traits' about individuals. 'Computer vision' techniques 'make sense' of image or video data by measuring, using machine learning to transform those measurements into 'representations', and subsequently into decisions and classifications. The 'online profiling' discussed over the past several decades, typically concerned knowledge production from 'transaction generated information' created by interacting with informational environments,[3] framing individuals in terms of proxies and scores.[4] Computer vision (and other biometric) techniques however, use extremely granular measurements of 'the real world' and increasingly deploy a logic of *exposing* or *revealing* the truth of reality and people within it. This article explores the nature of those knowledge claims and what might be the appropriate register of legal intervention to address data science's dominance as a paradigm for knowing people.

---

[*] Dr Jake Goldenfein is a postdoctoral research fellow at the Digital Life Initiative, Cornell Tech, Cornell University and Lecturer in Law at Swinburne University of Technology.
[1] See e.g. Isak Mendoza and Lee Bygrave, 'The Right not to be subject to Automated Decisions based on Profiling' in Taitani Synodinou et al (eds) *EU Internet Law: Regulation and Enforcement* (Springer 2017); Dimitra Kamarinou, Christopher Millard, and Jatinder Singh, 'Machine Learning with Personal Data' (2016) *Queen Mary University of London School of Law Legal Studies Research Paper 247/2016*; Fred H. Cate et al, 'Machine Learning with Personal Data: Is Data Protection Law Smart Enough to Meet the Challenge?' (2017) *Articles by Maurer Faculty* 2634.
[2] *REGULATION (EU) No 2016/679 OF THE EUROPEAN PARLIAMENT AND OF THE COUNCIL of 27 April 2016 on the protection of natural persons with regard to the processing of personal data and on the free movement of such data, and repealing Directive 95/46/EC (General Data Protection Regulation). OJ L 119 (2016-05-04), 1–88,* Article 4.
[3] Oscar H Gandy Jr, 'Statistical Surveillance: Remote Sensing in the Digital Age' in Kristie Ball, Kevin D Haggerty and David Lyon (eds) *Routledge Handbook of Surveillance Studies* (Routledge 2012).
[4] See e.g. Bernard Harcourt, *Exposed: Desire and Disobedience in the Digital Age* (Harvard University Press, 2015) discussing 'doppelganger logic', as well as Danielle Keats Citron and Frank Pasquale, 'The Scored Society: Due Process for Automated Predictions' (2014) 89 *Washington law Review* 1.

Computer vision generates classifications and knowledge about scenes, objects, events and people. Computer vision is not the only domain of data science articulating this new epistemological stance. Similar ideas are evident in numerous data science applications, often those involving biometrics. Computer vision does however, provides a useful entry point into what is qualitatively new about the empirical and epistemological position being developed. When applied to people, computer vision has clear legal and social significance. Facial recognition for instance, one such notorious application, links a person's portrait to some form of institutional identity like a driver's licence.[5] By connecting an individual to their behaviour in physical space, facial recognition couples identity across physical and computational registers, enabling automated decisions to be articulated into physical environments.[6] Both the intrinsic limitations and problematic applications of these technologies are well documented,[7] with calls for regulation not far behind.[8] However, there is already much more that computer vision can do when 'looking at people'.[9] Beyond 'identification', computer vision can also 'classify' people.[10] As well as linking your image and spatial location to an institutional identity, computer vision can make decisions about non-visual attributes such as what type of person you are, what you are doing, what you are feeling, and how you are likely to act in the future. These decisions may be about identity, gender, emotional state or future behaviours. But more controversial classifiers have also addressed questions of sexuality, criminal propensity, political position, IQ, workplace suitability, and paedophilic tendencies.[11] That type of 'computational physiognomy' is one element of an emerging field of 'personality computation', often described as Apparent Personality Analysis (**APA**) or Apparent Personality Recognition (**APR**).[12]

---

[5] See e.g. Jordan G Teicher, 'What Do Facial Recognition Technologies Mean for our Privacy?' *New York Times* (18 July 2018) <https://www.nytimes.com/2018/07/18/lens/what-do-facial-recognition-technologies-mean-for-our-privacy.html>; Clare Garvie, Alvaro Bedoya, and Jonathan Frankle, 'The Perpetual Lineup: Unregulated Police Face Recognition in America' (18 October 2018) <https://www.perpetuallineup.org/>; Monique Mann and Marcus Smith, 'Automated Facial Recognition Technology: Recent Developments and Approaches to Oversight' (2017) 40(1) *University of New South Wales Law Journal* 121.
[6] A similar process happens with vehicle registration and automated number plate recognition. See e.g. James Bridle, 'How Britain Exported Next-Generation Surveillance' (18 December 2013) <https://medium.com/matter/how-britain-exported-next-generation-surveillance-d15b5801b79e>.
[7] See e.g. Robin Kramer and Ray Ritchie, 'The Trouble with Facial Recognition technology (in the real world)', *The Conversation* (14 December 2016); Mike Orcutt, 'Are Face Recognition Systems Accurate? Depends on Your Race); *MIT Technology Review* (6 July 2016) <https://www.technologyreview.com/s/601786/are-face-recognition-systems-accurate-depends-on-your-race/>.
[8] See e.g. Jonathan Vanian, 'Microsoft President: Facial Recognition Technology Needs Government Regulation' *Fortune* (13 July 2018) <http://fortune.com/2018/07/13/microsoft-facial-recognition-regulation/>.
[9] The term 'looking at people' is an umbrella for research on computer vision applied to persons, see e.g. Sergio Escalera et al, 'ChaLearn Looking at People: A Review of Events and Resources' *Proceedings of the 2017 International Joint Conference on Neural Networks* (IJCNN 2017), IEEE, 2017.
[10] Gandy, above n 3.

Personality computation uses faces, postures, movements, actions, gestures, interactions, and emotions (as well as whether those emotions and expressions are real or fake) to infer personality traits.[13] Different analytic techniques operate on different data inputs. Some use dynamic information such as what a person *does,* or the way a physical morphology changes.[14] Some use static information such as how a person *looks*. Some use multiple modalities – like combinations of visual and audio data in different configurations. Whatever the particular decision or classification, the premise of these techniques is generating knowledge by applying data science to images in a way that transforms *measurements* of the world in the form of pixel vectors into classifications.

The scientific merit of computer vision profiling, especially when applied to non-visual characteristics of persons, has been rigorously challenged.[15] But the scientific validity of these projects is not the focus of this article. As Ian Hacking reminds us, there is a long history of producing social knowledge through new systems of measurement, both at the level of groups and individuals, and that 'new classifications and new enumerations are inseparable'.[16] The purpose of this article is therefore to examine those systems of enumeration and classification, try and understand the basis on which they produce knowledge about people, and explain why this presents unique challenges. The profiling exercises outlined here are thus presented as examples of the shift towards *computational empiricism* or *positivism* as a dominant knowledge system likely to have ongoing effects in juridical and political practices. While the knowledge claims of computer vision profiling are often modest or tentative, in certain papers authors have begun to argue that *only through computation* are we able to access all the information available in human faces and truly know people. In other words, only through computation can the excesses of the physical world and the people within it be perceived, processed and understood.

---

[11] Yilun Wang and Michal Kosinski, 'Deep Neural Networks are More Accurate than Humans at Detecting Sexual Orientation from Facial Images' (2018) 114(2) *Innovations in Social Psychology* 246; See e.g. https://www.faception.com/.

[12] Julio C.S. Jacques Junior et al, 'First Impressions: A Survey on Computer Vision-Based Apparent Personality Trait Analysis', arXiv:1804.08046v1 [cs.CV] (21 Apr 2018).

[13] Ibid.

[14] See e.g. Anna Hoffman and Luke Stark, 'Hard Feelings – Inside Out, Silicon Valley, and Why Technologizing Emotion and Memory is a Dangerous Idea' (11 September 2015) *LA Review of Books*, available <https://lareviewofbooks.org/article/hard-feelings-inside-out-silicon-valley-and-why-technologizing-emotion-and-memory-is-a-dangerous-idea/> where they discuss the problems of Paul Elkman's 'Facial Action Coding System', which measured the way people's faces change when they express emotions, suggesting the use of *dynamic* information does not necessarily solve any problems.

[15] See e.g. Blaise Agüera y Arcas, Margaret Mitchell, and Alexander Todorov, 'Physiognomy's new Clothes' (7 May 2017) <https://medium.com/@blaisea/physiognomys-new-clothes-f2d4b59fdd6a>; Frank Pasquale, 'When Machine Learning is Facially Invalid' (2018) 61(9) *Communications of the ACM* 24.

[16] Ian Hacking, *The Taming of Chance* (Cambridge University Press 1990), 53.

While the trope that technology facilitates greater knowledge of nature is a familiar one, it still requires close analysis in the specific context of computer vision. That analysis has become more imperative in the context of Trevor Paglen's identification that 'Human visual culture has become a special case of vision, and exception to the rule' in a situation where actually 'The overwhelming majority of images are now made by machines for other machines'.[17] While computer vision 'images' may be increasingly 'invisible', they still operate within the dynamics of representation and knowledge, and they still interact with human visuality. Western thought has spent a good deal of time reflecting on 'the connection between actuality and photo',[18] as well as the consequences of living in a world coded by undecipherable 'technical images'.[19] This material offers a useful touchstone for understanding the computer vision project of 'looking at people', and its political and epistemological consequences. While there has been a great deal of exceptional work on the problematic application of computer vision technologies, the goal of this article is thus less about exposing bias or lack of fairness, and more about understanding the 'programs' of the computer vision 'apparatus'.[20] Rather than challenge computer vision at the level of problematic applications, this article instead argues that, to avoid becoming a vector of data science's politics of completion, legal narratives should highlight the fundamental contingency of computational empiricism and its knowledge claims.

---

[17] Trevor Paglen, 'Invisible Images (Your Pictures are Looking at You)' *The New Inquiry* (8 December 2016) <https://thenewinquiry.com/invisible-images-your-pictures-are-looking-at-you/>.

[18] Walter Benjamin, 'A Short History of Photography 1934' in Alan Trachtenberg (ed) *Classic Essays on Photography* (Leete Island Books 1980).

[19] Vilém Flusser *Towards a Philosophy of Photography* (Anthony Mathews trans in Reaktion Books 2000) originally published as *Für Eine Philosophie der Fotographie* (European Photography 1983).

[20] Ibid, these terms are intended to follow the definitions provided by Vilém Flusser, the book includes a Glossary: At (83) '*Apparatus* (pi. *-es):* a plaything or game that simulates thought *[trans.* An overarching term for a non-human agency, e.g. the camera, the computer and the 'apparatus' of the State or of the market];

This article begins with a historical and technological contextualisation of personality computation. It looks at similar exercises from the past, and tracks the history of combining photographic portraiture with statistical analyses. That section demonstrates the ongoing ideologies behind this type of profiling, and how they are serviced by 'big data epistemology'. However, this is identified as only one of the political problems of this type of classification – another being the reverence for the forms of empiricism operationalised. Through the historical analysis, the idea is to demonstrate not how statistics and mathematics are becoming instruments of real or symbolic violence, but rather the nature of the change from visual to statistical forms of knowing. The article then seeks to sketch an outline of this emerging computational empiricism by connecting the knowledge claims of computer vision profiling to antecedent media technologies. The goal is to understand the ongoing role of 'representation' within the systems of measurement at work in computer vision, and the politics of their dissimulation. Finally, the article explores possibilities for legal and technical intervention. In that section, it is argued that the narratives animating existing legal protections are becoming deficient in the face of technological practices that see humans in a new way, primarily as patterns of information. Accordingly, the article suggests that when dealing with profiling, privacy and data protection now need to focus as much on 'metaphysics' as they do on 'metadata'.[21] Rather than the elimination of bias or unfairness, or the total prohibition of certain applications (both of which have their place within legal armatures), the argument here is for legal mechanisms that marginalize purely computational ways of knowing. The goal is not to 'put the genie back in the bottle' but rather highlight and contest the ontological power of computer vision when classifying people.[22]

---

organization or system that enables something to function.' At (84) '*Program: a* combination game with clear and distinct elements *[trans.* A term whose associations include computer programs, hence the US spelling].'

[21] Dan McQuillan, 'Data Science as Machinic Neoplatanism' (2017) 31(2) *Philosophy & Technology* 253, 263 where he notes 'With data science, we have moved from metadata to metaphysics; it is an embedded, even weaponised, philosophy.'

[22] Rob Kitchin and Martin Dodge, *Code/Space: Software and Everyday Life* (MIT 2011), 26.

**Computer Vision Profiling**

Computer vision is a family of technologies and practices by which computational systems come to understand and make decisions about the physical world and the people within it. When 'looking at people', applications range from relatively benign to highly problematic. The more problematic applications do, however, typically reveal more in terms of embedded technical, ideological, and epistemological assumptions. Forms of computational analytics have been used in personality analysis since the 1990s. But the history of combining photography with statistics to explore correlations between external appearance and internal personality goes back much further. Kate Crawford reminds us that, 'If algorithms present us with a new knowledge logic, then it is important to consider the contours of that logic and by which histories and philosophies it is most strongly shaped.'[23] Connecting computer vision projects to previous photographic ones thus demonstrates what lurks behind these systems of human classification, and how previous claims about the 'mechanical objectivity' of the camera have been displaced into the computational objectivity of statistics. But rather than demonstrate how data science can be a vector for ideology, the goal is to demonstrate what else is at stake in the computer vision project when 'looking at people' – in particular the proliferation of computational empiricism, the difference between visual and statistical knowledge, and the idea that data science can know people better than they know themselves.

*Classification and Ideology*

Criminal portraits were one of the first datasets systematically produced after the invention of the daguerreotype photographic process.[24] Founder of British Eugenics, bio-metrician, and Charles Darwin's cousin, Sir Francis Galton used these portraits to create 'composite' images intended to expose the 'mean' appearance of criminality.[25] Both phrenology and physiognomy had already been scientifically discredited by the time of Galton's experiments in 1877. However, belief in photography's 'mechanical objectivity' gave Galton a new (though crude) statistical tool to investigate his hypothesis of biological degeneration.[26] Galton's experiments were ultimately failures. He found that the *visual* similarities of the criminal classes disappeared through the composite process which instead revealed the 'common humanity in all'.[27] Despite this failure however, and despite the stigma associated with eugenics and 'Social Darwinism', the experiment has been repeated using new photographic techniques, statistical methodologies, and biological theories. One important differentiation however, is the movement in knowledge paradigm from qualitative visual searches for similarity amongst groups towards purely quantitative statistical analysis.

---

[23] Kate Crawford, 'Can an Algorithm be Agonistic? Ten Scenes from Life in Calculated Publics' (2016) 41(1) *Science, Technology & Human Values* 77, 85-6.
[24] See e.g. Jens Jäger, 'Photography: a means of surveillance? Judicial Photography 1850 – 1900' (2001) 5(1) *Crime, History & Societies* 27.
[25] Francis Galton, 'Composite Portraits, Made by Combining Those of Many Different Persons into a Single Resultant Figure' (1879) 3 *Journal of the Anthropological Institute of Great Britain* 132.

In parallel with ongoing physiognomic experimentation, there has been a growing psychological literature exploring how 'first impressions' are generated from looking at faces (including impressions of criminality).[28] That material demonstrates how faces *are* a form of non-verbal communication and part of the intrinsic heuristics we use to navigate daily social interactions. However, it has also led to the claim that 'research on appearance-based [personality assessment] might seem more credible if society were "not so enamoured of the idea that because a person's appearance *ought* not to make a difference, it *does* not," blaming the dearth of this research on the "naturalistic fallacy" (that is, confusing how things *are* with how they *ought* to be).'[29] In rejecting the 'naturalistic fallacy', the theoretical justifications for this work have at least moved away from the biological or genetic determinisms of Social

---

[26] Lorraine Daston and Peter Galison explain the term 'mechanical objectivity' in their text *Objectivity* (MIT Press, 2007), 121 as 'the insistent drive to repress the wilful intervention of the artist-author, and to put in its stead a set of procedures that would, as it were, move nature to the page through a strict protocol, if not automatically'. They note how the camera could 'quiet the observer so nature could be heard', and '"[l]et nature speak for itself"'.

[27] Galton, above n 25, 135.

[28] See e.g. J Willis and A Todorov, 'First Impressions: Making up your mind after a 100ms exposure to a face' (2006) 17(7) *Psychological Science* 592; Nalini Ambady and John Skowronski (eds) *First Impressions* (The Guilford Press 2008); Christopher Y Olivola and Alexander Todorov, 'Fooled by First Impressions? Re-examining the diagnostic value of appearance-based inference' (2010) 46(2) *Journal of Experimental Social Psychology* 315.

[29] Jeffrey M Valla, Stephen J Ceci, Wendy M Williams, 'The Accuracy of Inferences about Criminality Based on Facial Appearance' (2011) 5(2) *Journal of Social, Evolutionary, and Cultural Psychology* 66, 68 (quoting E. Berschied, 'An overview of the psychological effects of physical attractiveness' in G. Lucker et al *Psychological Aspects of Facial Form* (University of Michigan Press 1981)).

Darwinism, and is justified with other ideas.[30] Some explicitly avoid the politics of investigating correspondences between personality and appearance by claiming their work interrogates the pseudoscience of physiognomy in order to understand human subjective judgments (i.e. stereotyping). Others deploy Darwin's finding of adaptive significance in the capacity to make quick social judgments – what is called 'environment of evolutionary adaptation studies'[31] – to provide an 'ecological' rather than 'biological' theory. On that basis, researchers like Valla, Ceci and Williams argue that 'faces, even in the absence of dynamic behaviour, may be informative'.[32] In their study of the accuracy of physiognomic interpretation, they cite multiple studies showing that accurate physiognomic judgments can be made by humans.[33] Whether it is genetic determinism, biased social information processing, or 'ecological' accounts, there has however, never emerged a stable and accepted physiognomic theory. More than the success of 'evolutionary adaption' theories, it is the absence of an accepted or unified theoretical justification that has become the mechanism by which these physiognomic experiments continue. Although the relationship between causation and correlation in scientific method is complex,[34] these projects seem content, in Kuhn's terms, to travel the road from scientific law to measurement in reverse,[35] and focus on correlations entirely unpinned from theoretical accountability. To that end, explorations of facial morphology through statistical methods have proliferated and gained disciplinary legitimacy, with entire fields and competitions (sponsored by industry actors like Microsoft,

---

[30] See e.g. Alexander Todorov, *Face Value* (Princeton University Press 2017); Thomas Alley, 'Physiognomy and Social Perception' in Thomas Alley (ed) *Social and Applied Aspects of Perceiving Faces* (Laurence Erlbaum Publishers 1988), where he neatly categorises common approaches in physiognomy including: (a) seeking resemblances in appearance between humans and animals, assuming similarities in appearance indicate shared psychological qualities (e.g., a person whose face is reminiscent of a fox is believed to be sly and cunning); (b) an inductive process whereby others sharing distinctive facial characteristics with a known person (or group) are believed to share some of the familiar person's (group's) psychological characteristics; (c) an approach based on functional analogies whereby inferences are made according to the size, shape, or presence of facial features and their common function(s); and (d) an approach based on the facial expressions engendered by various emotional and cognitive states that looks for *traces* of these expressive facial postures as clues to the common emotional and cognitive states of an individual.
[31] Valla et al, above n 29.
[32] Ibid, 69.
[33] One example is a 1962 study on inferences of criminality that claimed success in replicating Galton's process. Kozeny's 'Experimental investigation of physiognomy utilizing a photographic-statistical method' used 730 criminal portraits to create composites for 16 crime categories, which were then physionogmically classified (from the visual information) in E Kozeny 'Experimentelle Untersuchungen zur Ausdruckskunde mittel photographisch-statisticher Methode' (1962) 114 *Archive für die gesamte psychologie* 55.
[34] See e.g. Roddam Narasimha, 'Axiomatism and Computational Positivism: Two Mathematical Cultures in Pursuit of Exact Sciences' (2003) 38(35) *Economic and Political Weekly* 3650.
[35] Thomas Kuhn, 'The Function of Measurement in Modern Physical Science' (1961) 52(2) *Isis* 161, 189-190.

Facebook, Amazon and even Disney) for personality and social interaction typing through image and video data. With its capacity to surpass the visual with the quantitative, computer vision represents the technological boon that Galton was missing.

*Computational Personality Analysis*

The success of machine learning and particularly convolutional neural networks in computer vision mean APA and APR have become significant sub-fields in computer vision research. Whatever the specific technique, this research assumes a stable statistical relationships between a stimuli and social perception of personality (in APA), or true personality characteristics (for APR).[36] Whereas APA avoids interrogating the accuracy of its claims, instead focusing on personality traits attributed by others (often Amazon Turk workers), APR on the other hand, seeks to identify the true personality of the individual. Although these disciplines explore multiple stimuli including *dynamic* facial information, handwriting, and speech, they also include what could be called 'computational physiognomy'. Early work on computational physiognomy through computer vision was based on very simplistic models using Euclidean distances (the straight-line measurement between 2 points in a geometric plane) and geometric angles to classify facial features corresponding with the pioneering physiognomer Lavator's division of the face into 32 classes.[37] Initially, these experiments were framed in terms of investigating whether computers could replicate the trait evaluation performed by humans (i.e. first impressions analysis). There were no complex machine learning methods, or assessments of accuracy. It was simply a translation of the task of physiognomic measurement into a computer vision system. From that point however, this type of personality computation became far more sophisticated.[38] It was not long until convolutional neural networks were predicting intelligence and other personality characteristics based on images.[39] Competitions for computational personality analysis (both apparent and true) began in 2012, with 'deep learning' dramatically improving the early unimpressive results in classifier accuracy.[40] Deep learning enables 'measurement' at a radically different scale than earlier classifiers that used engineered low-level image 'features' (like facial landmarks) or intermediate level 'attributes' (like nose, glasses, moustache, hair, and forehead height).

---

[36] Jacques Junior et al, above n 13.
[37] Hee-Deok Yang and Seong-Wang Lee, 'Automatic Physiognomic Analysis by Classifying Facial Component Feature' (2006) *IEEE The 18th International Conference on Pattern Recognition (ICPR'06)* – although they reference work statistical work from 1997 in V Dominique et al, 'What Represents a Face? A Computational Approach for the Integration of Physiological and Psychological Data' (1997) 26 *Perception* 1271. And material on computer vision for expression analysis has also been continuing since the 1990s; See e.g. John Graham, 'Lavater's "Physiognomy": A Checklist' (1961) 55(4) *The Papers of the Bibliographical Society of America* 297. This pioneering work was initially published in 1806.
[38] These advances were typically based on research into object detection using convolutional neural networks in A Krizhevsky et al, 'ImageNet Classification with deep convolutional neural networks' (2012) 25 *Proceedings of Advances in Neural Information Processing Systems* 1090.

How neural networks change the basis of knowledge production as compared to visual systems is demonstrated very clearly in several computer vision and data science projects, for instance Wu and Zhang's 2016 paper 'Automated Inference on Criminality Using Face Images'.[41] That paper used 1856 government ID images, about half of which included individuals with a criminal conviction, as a dataset for a criminal-propensity classifier.[42] The theoretical grounding of this research is similar to other APR exercises, ostensibly testing the social inference hypothesis.[43] But the research is more revealing in terms of how computer vision systems claim to generate knowledge about individuals. Most telling is the authors' acknowledgment that the variance between criminal and non-criminal populations is not evident from visual assessments or simple Euclidean measurements.[44] In other words, visual information, in the sense of information that is visually perceivable and interpretable, what Galton had unsuccessfully relied on, was insufficient for physiognomic purposes. Only through the higher dimensional computational analysis of numerical quantitative measurements was this statistical separation discernible. This finding represents the premise and utility of neural networks for this application,[45] and is highly illustrative of the shift from qualitative visual to quantitative statistical ways of knowing people. This change in the nature of knowledge production is made even more explicit in Wang and Kosinski's controversial physiognomic work 'Deep Neural Networks are More Accurate than Humans at Detecting Sexual Orientation from Facial Images'.[46]

---

[39] Ting Zhang et al, 'Physiognomy: Personality Traits Prediction by Learning' (2017) 14(4) *International Journal of Automation and Computing* 386. Work of this type often uses the OCEAN model for personality description.
[40] Jacques Junior et al, above n 13, discussing the Interspeech, MAPTRAITS, WCPR and ChaLearn competitions.
[41] Xiaolin Wu and Xi Zhang, 'Automated Inference on Criminality using Face Images' arXiv:1611.04135v1 [cs.CV] (13 Nov 2016).
[42] These authors built a criminality classifier not on the basis of a statistical mean as sought by Galton, but by measuring variance from a non-criminal normal (i.e. based on the idea that there is a greater resemblance in non-criminal faces).
[43] Ibid 2.
[44] Ibid 6.
[45] Yann LeCun et al, 'Deep Learning' (2015) 521 *Nature* 436.

That project used 14438 facial images extracted from an existing database, of which 6076 were identifiably gay according to correlation with a Facebook Audience Insights platform, and self-declared sexual interest on Facebook. A deep neural network called VGG-Face (originally deployed for facial recognition) that transforms facial images into 4096 particular scores, was used for pattern analysis and classification.[47] While the authors begin from the proposition that 'DNNs are increasingly outperforming humans in visual tasks such as image classification, facial recognition, or diagnosing skin cancer',[48] they ultimately conclude that 'our faces contain more information about sexual orientation than can be perceived or interpreted by the human brain'.[49] This finding suggests that if the data sets can be constructed, and the results are to be accepted, there are few limitations to what characteristics might be inferred from images.

There is something historically significant in physiognomy being a platform for the advancement of computational empiricism. Criminologist Nicole Rafter has brilliantly argued with respect to the earlier (analogue) practices of physiognomy and phrenology that their scientific invalidity was insignificant when compared to the epistemological paradigm that they ushered in.[50] Those pseudo-sciences shifted how we understand people and their behaviour away from metaphysics and theology and towards 'analytical empiricism'. Rafter shows how phrenology, the discredited science of 'the correspondences between the external and internal man, the visible superficies and the invisible contents'[51] but based on reading character traits from skull morphology, therefore 'produced one of the most radical reorientations in ideas about crime and punishment ever proposed in the Western world.'[52] It was radical for its providing a *measurable* explanation of criminal behaviour and other social phenomena on the basis of positivist reasoning with empirical methods, and it transformed criminology, penology, and jurisprudence by changing how we generate knowledge about people. It is argued that the profiling techniques described above should be thought of as data points in the movement towards *computational empiricism* as a dominant knowledge system. Beyond 'laundering' various ideological, even eugenicist, politics through technological neutrality,[53] the claim here is that the foundational knowledge claims of these technical systems exhibit their own political, even philosophical, position.

---

[46] Wang and Kosinski, above n 12.
[47] The results were post-hoc justified with a prenatal hormone theory (that over or under exposure of androgens during gestation affects both facial appearance and sexual orientation), which gave a biological foundation to the statistical findings.
[48] Ibid 247.
[49] Ibid 254.
[50] Nicole Rafter, 'The Murderous Dutch Fiddler: Criminology, History and the Problem of Phrenology' (2005) 9(1) *Theoretical Criminology* 65.
[51] Ibid at 71 quoting JC Lavater, 'Essays on Physigonomy' Abridged from Mr Holcroft's translation. London: Printed for G.G.J. & J. Robinson.

Computational physiognomy, like its analogue predecessor, is not the only application of this empirical paradigm. Rather, it is best understood as a harbinger of an evolving epistemological environment. Other biometric data science applications operate on a similar knowledge-logic. For instance, automated analysis of voice and speech are being used to assess ethic origins for migration assessments,[54] as well as making customs assessments of migrants and tourists.[55] There is also a growing health-tech industry, where recording of particular bodily functions like sleeping and eating, speaking, or physical interactions with a computer are being computed into assessments of physical and psychological health.[56] The accuracy of these systems is not going to determine their epistemological validity. Projects are being funded, and the political will is there to deploy them. The empirical sciences accordingly follow, building new ways to classify and know persons. Data science, and its particular systems of measurement and classification, are thus becoming the prime-movers in building this new epistemological terrain.

---

[52] Ibid 65.
[53] Agüera y Arcas et al above n 15.
[54] Emily Apter, 'Shibboleth: Policing by Ear and Forensic Listening in Projects by Lawrence Abu Hamdan' (2016) 156 *October* 100.
[55] Douglas Heaven, 'An AI lie detector will interrogate travellers at some EU borders', (31 October 2018) *NewScientist* <https://www.newscientist.com/article/mg24032023-400-an-ai-lie-detector-will-interrogate-travellers-at-some-eu-borders/>
[56] See e.g. Saeed Abdullah and Tanzeem Choudhury, 'Sensing Technologies for Monitoring Serious Mental Illnesses' (2018) 25(1) *IEEE MultiMedia* 61; Rui Wang et al, 'Predicting Symptom Trajectories of Schizophrenia using Mobile Sensing' (2017) 1(3) *Proceedings of the ACM on Interactive, Mobile, Wearable and Ubiquitous Technologies* 110.

**Computational Empiricism**

The claim that nature can only be understood through computation articulates data science as an attempt to access the hidden mathematical sub-structures of reality. In many respects, this claim is neither unique nor controversial, but it remains significant in the context of a technological capacity that is proliferating so rapidly and widely. It is also far from universally accepted. Some computer scientists sensibly argue that 'deep learning can't extract information that isn't there, and we should be suspicious of claims that it can reliably extract hidden meaning from images that eludes human judges.'[57] But such admissions seem peculiar when, really, deriving meaning from measurements too granular for non-computational analysis is the very premise of machine learning. Further, as argued, the accuracy of those applications is unlikely to be the primary determinant of whether their outputs are interpreted as "true". To that end, some have begun to argue that claiming access to the hidden structures of reality has now become the organizing principle of data science.[58]

Many technologies of representation claim to reveal previously unseen information. Photography, for instance, was understood as a way to 'see into nature's cabinet'.[59] It challenged the 'optical unconscious'.[60] Walter Benjamin describes this access to nature as opening up 'in this material the physiognomic aspects of the world of images, which reside in the smallest details, clear and yet hidden enough to have found shelter in daydreams.'[61] Other optical technologies offer similar narratives. The telescope gave access to celestial knowledge – the microscope to cellular knowledge. X-ray imaging and radiographic measurements of material density afforded a 'New Sight'[62] that could reveal 'hidden existence'.[63] Roberta McGrath describes public attitudes around x-rays, noting 'The body itself is thus perceived as literally ghost-like, immaterial, only flesh, as being merely a thin veneer, literally a skin, covering a hidden, deeper reality which will, like truth, be uncovered, revealed.'[64] Artists and intellectuals similarly responded, 'Who can still believe in the opacity of bodies, since our sharpened and multiplied sensibilities have already penetrated the obscure manifestation of mediums?'[65] The idea of the body as readable medium and material manifestation of a deeper informational truth is also highlighted in Mark Andrejevic's analysis of 'neuromarketing', where he describes how: 'The notion that bodies are, for marketing purposes, more truthful than the words they utter is emerging as a recurring theme in the promotion of neuromarketing, which promises to render obsolete the allegedly quaint and outdated techniques of surveys and focus groups… They can, thanks to new technologies, cut through directly to the underlying truths revealed by the brain.'[66]

---

[57] Agüera y Arcas et al above n 15, although those authors do then give the conflicting argument that 'On a scientific level, machine learning can give us an unprecedented window into nature and human behaviour, allowing us to introspect and systematically analyze patterns that used to be in the domain of intuition or folk wisdom.'
[58] McQuillan, above n 21.
[59] Roberta McGrath, *Seeing her sex: Medical archives and the female body* (Manchester University Press 2002) 114.
[60] Benjamin, above n 18.
[61] Ibid.
[62] 'The New Sight and the New Photography', *The Photogram* (1898).
[63] Mary Warner Marien, *Photography: A Cultural History* (2nd ed) (Laurence King Publishing 2006), 216.

Andrejevic places this claim within 'The appeal of techniques for bypassing discursive forms of representation (by cutting 'straight to the brain')' against 'the popularised and reflexive mediated critiques of discursive forms of representation for their potentially deceptive, indeterminate, and constructed character.'[67] These approaches can thus be genealogically situated amongst the behavioural sciences, wherein the premise of stable relationship between stimuli and character afforded mechanisms for personality and behaviour computation such as 'digital phenotyping'.[68] They also represent an historical trajectory of understanding human subjectivity as an information pattern,[69] bound within the material container of the body that becomes its expression. As William Gibson said, 'data made flesh'.[70] Computer vision and biometric data science add another dimension to these narratives however. Photographs, telescopic observations, radiographic photograms, and even neuromarketing enabled a form of discovery that still requires human interpretation. In computer vision, the measurement, encoding and decoding, and knowledge discovery are increasingly automated. Through machine learning and neural networks, even the selection of representations – that is, the mechanisms by which the world is presented for analysis – is also automated. The elements of images from which meaning is derived, often called 'features', are selected through automated learning processes instead of the laborious manual process of 'feature engineering'. Through these processes, the clinical is replaced with the empirical, image comprehension is replaced by image computation, observation replaced with measurement, with the truth – the deeper hidden reality – only becoming available in the high-dimensionality that neural networks can process.[71] This is not the combination of inputs from 'multiple large data sources to generate new hypotheses about the way the world works and prescriptions for how to act upon that knowledge'[72] typically associated with big data and predictive analytics. It involves a deeper belief in how the world *can* be understood, premised on new systems of measurement, new systems of representation, new understandings of human subjectivity, and new forms of statistical analysis.

---

[64] McGrath, above n 59, 115.
[65] Umberto Boccioni, quoted in Didier Ottinger (ed) *Futurism* (Centre Pompidou / 5 Continent Editions 2008), 154.
[66] Mark Andrejevic, 'Brain Whisperers: Cutting Through the Clutter with Neuromarketing' (2012) 2(2) *Somatechnics* 198, 199.
[67] Ibid 200.
[68] See e.g. Luke Stark, 'Algorithmic Psychometrics and the Scalable Subject', (2018) 42(2) *Social Studies of Science* 204.
[69] See e.g. Paul Edwards, *The Closed World: Computers and the Politics of Discourse in Cold War America* (MIT Press 1996); N. Katherine Hayles, *How We Became Post-Human: Virtual Bodies in Cybernetics, Literature and Infomatics* (University of Chicago Press 1999).
[70] William Gibson, *Neuromancer* (Ace 1984).

*Measurement and Representation*

Accessing the hidden mathematical substructure of reality requires numbers. Reducing the real world to numbers requires measurement. Accordingly, computer vision systems do not see – they measure. They measure visual data (**x**) in order to determine a 'world state' (**w**).[73] As Sheila Jasanoff notes, 'Any form of data collection involves, to begin with, an act of seeing and recording something that was previously hidden and possibly nameless'.[74] In the same way that photography bypassed the optical unconscious, computer vision profiling is about noticing, measuring, and analysing that which was previously not available to human perception and cognition. The process is complex however because the relationship between **x** and **w** is not one-to-one. Photographic images are reductions of the three-dimensional world into a two-dimensional set of measurements. There are however multiple configurations of the three-dimensional world that might result in any particular two-dimensional output. In other words, 'there may be many real-world configurations that are compatible with the same measurements.'[75] Thus the chance that any possible world state is present or true is described using probability. The probabilistic techniques of statistical clustering used to generate meaning about the world state typically use machine learning. But machine learning is computationally demanding, and it cannot operate on absolutely all the data that could be derived from an image. Instead, it is necessary to marshal selections of data into 'representations'.

---

[71] Rob Kitchen, 'Big data, New Epistemologies and Paradigm Shifts', (2014) *Big Data and Society* 1, at 2 where he says, 'Big Data analytics enables an entirely new epistemological approach for making sense of the world; rather than testing a theory by analysing relevant data, new data analytics seeks to gain insights "born from the data".

[72] Sheila Jasanoff, 'Vitual, Visible, and actionable: Data assemblages and the sightlines of justice' (2017) 4(2) *Big Data and Society* (online) <http://journals.sagepub.com/doi/full/10.1177/2053951717724477>.

[73] Simon JD Prince, *Computer Vision: Models, Learning and Inference* (Cambridge University Press 2012); Note Rob Kitchen, 'Big data, New Epistemologies and Paradigm Shifts', (2014) *Big Data and Society* 1 where he quotes Sinan Aral, who states 'Revolutions in science have often been preceded by revolutions in measurement.'

[74] Jasanoff, above n 72.

When computer vision emerged as a discipline in the 1960s, it became apparent that those systems could not respond to the totality of recorded signal (an image). Image data typically comes as RBG values per pixel inscribed by the translation (or transduction) of light energy to voltage in a sensor. The computational resources required to perform statistical analysis on that amount of data is immense. Therefore, instead of 'image level computation', computational economy requires transformations into symbolic representations.[76] The inability of computer vision systems to respond directly to the totality of registered signal has been acknowledged by some as a failure,[77] and others as a fundamental limitation.[78] Nevertheless, it remains the primary mechanism by which computer vision translates measurements of the world into the computational register. That means 'features' of visual data have to be selected for analysis, but as noted above that selection is increasingly automated. In the last few years significant research has gone into ways to avoid manual assessments of images to identify what might be relevant for building classifiers (i.e. feature engineering)[79] in favour of a statistical pattern recognition that can identify a relevant feature on the basis of its probabilistic relationship to other features in the environment.[80] This is sometimes called 'representation learning', where a system is fed with "raw data" 'to automatically discover the representations needed for detection or classification.'[81] 'Deep learning' means multiple layers of representation, automatically generated, where each level of representation is more complex and abstract than the previous one. As LeCun, Bengio and Hinton explain, 'The key aspect of deep learning is that these layers of features are not designed by human engineers: they are learned from data using a general-purpose learning procedure.'[82]

---

[75] Prince, above n 73.
[76] CM Brown, 'Computer Vision and Natural Constraints' (1984) 224 *Science* 1299.
[77] Prince, above n 73.
[78] Herbert Dreyfus, 'Why Heideggerian AI failed and how fixing it would require making it more Heideggarian' (2007) 171(18) *Artificial Intelligence* 1137, where he argues that the use of representations rather than direct sensory input is a limiting factor for artificial intelligence.
[79] Yoshua Bengio, Aaron Courville and Pascal Vincent, 'Representation Learning: A Review and New Perspectives' (2014) *arXiv:1206.5538v3 [cs.LG] 23 Apr 2014*.

The inherent paradox of this epistemological platform is the use of correlation to create invisible or inaccessible automated representations in order to surpass the use of manual representation. In other words, overcoming the limits of representation with more representations.[83] Framing it this way prompts us to reconsider imagistic representation within data scientific contexts. Much has been written about how knowledge is modulated through natural representations like the index and icon.[84] But now similar investigations are necessary for mathematical imagistic representations like 'functional representations' – where a function is fit to the discrete and finite set of measurements that constitute an image such as the pixel coordinates and values; 'linear representations' – where images are unwound into a vector matrices; 'spatial frequency representations' – that measure the speed at which a particular quality changes across an image surface; 'relational representations' – where images are represented with graphs; and 'probabilistic representations' – where mathematical tools are used to *estimate* the best version of a particular image given a measurement of a corrupted or noisy image.[85] Representation in computer vision has changed into a project of finding the target *thing* (i.e. the pattern) from all the information within an image. With real consequences for how we derive meaning from the visual world.

---

[80] See e.g. Hairong Qi and Wesley E. Snyder, *Machine Vision* (Cambridge University Press, 2004). They explain that when features extracted from an image most closely match features of a trained image, then a simple statistical (maximum likelihood) classifier is useful for classifying the content of an image. In other words, the maximum likelihood approach to classifier design uses statistical representations to describe the probability that an object having a certain set of measurement belongs to a particular class.
[81] LeCun et al, above n 45.
[82] Ibid 436.
[83] This is also discussed in Andrejevic, above n 66 at 212.
[84] See e.g. CS Peirce in Charles Hartshorne and Paul Weiss (eds), *Collected Papers of Charles Sanders Peirce* (Harland University Press, 1932) Volume 2, 281.
[85] Qi above n 80.

*Computational Empiricism as a Dominant Epistemology*

As society becomes more statistical,[86] it makes sense that both representation and knowledge take less deterministic forms. However, from the above, we can begin to draw an outline of new forms of computational empiricism (or positivism) as applied to understanding people.[87] First, it operates on the basis that external measurement or observation is a more reliable pathway to knowledge than the symbolic output of a subject. This is, of course, not unique to these practices or technologies, it is simply one element of the schema. It is also not a uniquely visual phenomenon. In addition to the techniques described above, the technology of the stethoscope and the practice of 'auscultation' – listening to the body at a physical distance – offers another useful example of technological mediation making the body an object of knowledge. Histories of diagnostic practices with stethoscopes have been invoked to demonstrate the movement from theoretical to perceptual ways of knowing the body, achieved through the combination of rationality and empiricism.[88] This line of thinking describes the ascendance of empiricism in this (medical) context as entwined with the construction of a new subject, and productive of a new medical epistemology of pathological anatomy.[89] The updating of remote sensing in medical applications with new sensors and data science is similarly oriented towards constructing epistemologies of pathological behaviours.[90] However, empirical data science engages with a different human subject from those older techniques, one that is simultaneously interpreted and constructed an informational pattern rather than embodied puzzle.

---

[86] Hacking, above n 16.
[87] Note the term 'computational empiricism' was used by Paul Humphries in 'Computational Empiricism' in Bas C. van Fraasen (ed) *Topics in the Foundation of Statistics* (1997, Springer) at 119, where he discussed the idea in the context of changing scientific methods associated with the adoption of powerful instrumentation and powerful computational devices. His point is to suggest notions of 'logical empiricism' require updating by computationally oriented methods. The term 'computational positivism' is best described in Narasimha, above n 34, where he uses the term to describe methodologies in exact sciences and mathematics that focus on matching algorithms to observations rather than drawing conclusions from axioms and models.
[88] See Jonathan Sterne, 'Mediate Auscultation, the Stethoscope, and the "Autopsy of the Living": Medicine's Acoustic Culture' (2001) 22(2) *Journal of Medical Humanities* 115, where he locates this in the work of Michelle Foucault's *The Birth of the Clinic* (1973, Pantheon Books), and S.J. Reiser's *Medicine and the reign of Technology* (1978, Cambridge).
[89] Ibid 117.
[90] See e.g. Saeed Abdullah and Tanzeem Choudhury, 'Sensing Technologies for Monitoring Serious Mental Illnesses' (2018) 25(1) *IEEE MultiMedia* 61; Rui Wang et al, 'Predicting Symptom Trajectories of Schizophrenia using Mobile Sensing' (2017) 1(3) *Proceedings of the ACM on Interactive, Mobile, Wearable and Ubiquitous Technologies* 110.

A second element to the schematic is a specific type of computational intervention in the relationship between measurement and classification. On one hand, computational systems differ from other representational technologies because of their capacity to both measure and process quantities of data that are too large for human tabulation, too discrete for human perception, and too complex for human cognitive analysis. This is also not unique to computer vision or data science. Where these techniques differ however, is the degree of automation in the selection of elements (or features or dimensions) of a measurement deemed to be meaningful. In other words, decisions about what elements of a measurement inform each layer of representation are increasingly displaced into automated learning systems. Those are highly political decisions about how the real world is translated into the symbolic register of computation, and are handed off to automated systems. Of course, humans participate in selecting input data (i.e. what the sensor captures) and defining the accuracy of outputs, which each participate in tuning the selection of representations and their parameters (i.e. weightings), but the process of producing representations is dissimulated into the architecture of the system.

A third element is a belief that this process is working towards exposing the fundamental substructures of reality. The belief that increasingly granular measurement and high dimensionality analysis has the capacity to reveal hidden truths is visible to a greater or lesser degree in research projects using high resolution sensors and learning applications. This computational variant of realism has been described by Dan McQuillan as 'machinic neoplatansim'.[91] That is, a metaphysical commitment to a world of truth, form, and idea existing behind, and only imperfectly imprinting on, the world of the humanly sensible, accessible only through mathematics. Rather than being a specific tool or method, for McQuillan, data science thus represents an automated and applied philosophy, maintaining an epistemological reverence for '[a] hidden layer of reality which is ontologically superior,

---

[91] McQuillan, above n 21, 261.

Electronic copy available at: https://ssrn.com/abstract=3284598

expressed mathematically and apprehended by going against direct experience.'[92] None of these observations are unique to technologies of computer vision, the practices of machine learning, or the disciplines of data science. However, when schematised and directed at understanding people, they represent an idiosyncratic system of knowledge production that challenges the way legal narratives (as well as many others) have been deployed to protect individuals in the context of online profiling.

**Law in the World State**

A classic critique of technology insists that technological mediation inhibits access to 'the real' or 'the event'. In the case of photography for instance, we are reminded that the images we create, while supposed to be windows or maps for understanding the world around us, actually operate more like 'screens'. Rather than expose the truth of the world, our images saturate the world, producing a veneer under which 'the real' slowly decays. A form of this critique is often levelled at digital profiling, wherein data produced through interactions with information environments are used as proxies for defining characteristics about us. That is the world of actuarial risk assessment and the 'scored society'. Critiques of those technologies describe how such 'scores' inadequately capture or represent individuals. Proxies result in reduction, distortion and error.[93] Justin Clemens and Adam Nash offer a useful account of this process wherein information 'must first be digitised to data, then modulated between storage and display in an endless protocol-based negotiation that both severs any link to the data's semantic source and creates an ever-growing excess of data weirdly related to, but ontologically distinct from, its originating data source.'[94] In other words, the mathematical codes and conventions used to analyse and parse already hyper-mediated digital information omit or marginalise their 'natural' starting point. Clemens and Nash thus claim that only through modulation into a display register does digital information obtain meaning.[95] This intervention frames the harm to persons from profiling in terms of *loss*. It also grounds, for instance, data protection's animating principle of 'transparency'. Legal narratives around transparency, particularly in the context of data protection, emerged to deal with that potential for *misrepresentation*, inaccuracy, or distortion in cases where individuals are transcribed into computational systems and categorised (i.e. profiled) on the basis of pattern analysis. Data subject rights of access and rectification are thus deployed to help individuals maintain the 'borderlines of meaning' about themselves as data is disclosed and processed across contexts.[96] However, as technical narratives change, they also challenge this legal mechanism.

---

[92] Ibid.
[93] See e.g. Cathy O'Neil, *Weapons of Math Destruction* (Crown Books 2016); Virginia Eubanks, *Automating Inequality* (St Martins Press 2018).
[94] Justin Clemens and Adam Nash, 'Being and Media: Digital Ontology after the Event of the End of Media' (2015) 24 *The Fibreculture Journal* 6, 9.
[95] Ibid 10.

The scores and proxies used in, for example, computer vision profiling are of a different class and scale than those used in actuarial analysis and other forms of data mining. The scores of 'behavioural computation' and 'personality computation' take the shape of a ledger, through which behavioural characteristics can be collated.[97] Through its new epistemological program, data science then starts to move from a logic of 'approximation' to a logic of 'revelation'. From the computational empiricist position in which the quantitative is the path to knowledge, the world of infinite possibility now exists in the data rather than in the human spirit. The 'natural' starting point is no longer omitted, but rather measured in a dimensionality that humans can neither access nor interpret. Profiling systems claiming deeper knowledge about persons thus challenge the utility of transparency as a legal narrative and dismiss the role of the captured subject in forming their own identity. Thinking through how law or any other system of governance might address these technologies accordingly means attending to this re-arrangement of technological practice and how it constructs persons.

Legal thinking has so far responded to these technologies and techniques in different ways. One register seeks improvement of automated systems' encoding of the world. The goal is to achieve a 'fairer' computational translation of the real world by exposing and limiting bias and prejudice. Unfair discrimination can find its way into automated systems in multiple ways,[98] and improving the outcomes of automated decision systems remains critically important. But scholars like Frank Pasquale have begun to ask whether this form of accountability adequately considers the question of 'accountability to whom'.[99] Without proper attention to that question, this form of accountability risks becoming part of the feedback mechanism that continually improves and thus proliferates automated decision making through legal optimisation.[100] Yarden Katz similarly comments that 'If AI runs society, then grievances with society's institutions can get reframed as questions of "algorithmic accountability." This move paves the way for AI experts and entrepreneurs to present themselves as the architects of society.'[101] Fairness and accountability through decision system optimisation thus fail to address the problem that computational empiricism should be understood as simply one way of knowing, especially when applied to persons. Rather than challenge the dominance of computational empiricism, 'fairness' projects focused exclusively on system improvement risk law becoming enrolled in data science's politics of completion. As Joshua Scannell notes, 'it is not at all clear that biometric accountability, accuracy, and fairness are mechanisms for achieving justice, or even a baseline common humanity. That is not what biometric systems do. Instead, they reflect the institutional demands of the entities that employ them — demands that are *always* intended to sort people into those with access or without; secured by or made insecure by the state; capacitated or incapacitated by a credit score.'[102]

---

[96] Lee Bygrave, 'Privacy and Data Protection in an international perspective', (2010) 56(8) *Scandinavian Studies in Law* 165.
[97] See e.g. Vlad Savov, 'Google's Selfish Ledger is an Unsettling Vision of Silicon Valley Social Engineering' (17 May 2018) *The Verge* available <https://www.theverge.com/2018/5/17/17344250/google-x-selfish-ledger-video-data-privacy>.

Another legal register challenges specific applications of data science on ethical or political grounds. This legal mode identifies where data science applications have too pernicious an effect on society according to liberal democratic values.[103] This is typically a more traditional 'privacy' mechanism defending the fundamental dignity and opacity of persons against overreaching applications. These projects similarly deploy their force at the level of application rather than at the epistemological or political bases of computational empiricism. Beyond abandoning or optimising the technologies however, there remains an important space for demonstrating the contingency of computational empiricism's knowledge claims and challenging data science's move from 'metadata' to 'metaphysics'.[104] This is what Mireille Hildebrandt, for instance, means when she talks of 'speaking law to the power of statistics'[105] – a program of re-inscribing uncertainty into automated knowledge production.

---

[98] Solon Borocas and Andrew Selbst, 'Big Data's Disparate Impact' (2016) *California Law Review* 104.
[99] Frank Pasquale 'Odd Numbers' in *Real Life Magazine* (20 Aug 2018), <http://reallifemag.com/odd-numbers/>.
[100] In other words, often highlighting bias participates in the positive feedback relationship that proliferates automated decision making and computational empiricism as knowledge systems – In Flusser, above n 19, this is identified as the program of the camera at (46). When bias is exposed in contemporary systems, industry moves to improve their decision making systems. See e.g. Dan Thorp-Lancaster, 'Microsoft improves facial recognition error rates across skin tones, gender' (26 June 2018) <https://www.windowscentral.com/microsoft-improves-facial-recognition-error-rates-across-skin-tones-gender>.
[101] Yarden Katz, 'Manufacturing an Artificial Intelligence Revolution' <https://papers.ssrn.com/sol3/papers.cfm?abstract_id=3078224>
[102] Joshua Scannell 'Controlled Measures' in *Real Life Magazine* (17 September 2018) <http://reallifemag.com/controlled-measures/>.

*Law and Computer Vision Profiling*

There are already important legal limitations on automated decision making and automated profiling in, for instance, the GDPR. Alongside well explored 'access and rectification rights',[106] and the fundamental principles of processing in Article 5,[107] Article 22 gives data subjects rights to not be subject to automated profiling decisions based solely on automated processing where those decisions produce legal or similar effects. There are numerous limitations to that latter provision, including explicit consent, enabling legislation, or satisfaction of a contract.[108] What constitutes legal or similar effects, as well as purely automated processing, are also unclear. That a human decision maker anywhere in the process might remove an automated system from the purview of the Article seems a problematic limitation, especially considering the growing evidence that human decision makers rarely contest the outcomes of decision support systems. Further, guidance by the Court of Justice of the European Union will be essential to a more complete understanding of

---

[103] See e.g. Evan Selinger and Woodrow Hartzog, 'Opinion: It's time for an about-face on facial recognition' (22 June 2015) *Christian Science Monitor* <https://www.csmonitor.com/World/Passcode/Passcode-Voices/2015/0622/Opinion-It-s-time-for-an-about-face-on-facial-recognition>.
[104] McQuillan, above n 21, 263.
[105] Mireille Hildebrandt, 'Law as Computation in the era of artificial legal intelligence: Speaking Law to the Power of Statistics' (2018) 68 *University of Toronto Law Journal* 12.
[106] See e.g. Arts 13, 14, 15 (access), 16 (rectification), 17 (erasure), 18 (restriction), 19 (notification), 20 (data portability),
[107] These include lawfulness, fairness, transparency, purpose limitation, data minimisation, accuracy, accountability, storage limitation, integrity and confidentiality.
[108] Art 22(2)(a), (b), (c).

Electronic copy available at: https://ssrn.com/abstract=3284598

these provisions. A right to 'explanation' has also been read into Art 15(1)(h), producing rigorous debate over what the Article truly affords data subjects, and how useful it may be.[109] While a similar (and arguably broader) provision has been in place since 1995, a broad reading may give data subjects useful mechanisms for not only understanding, but properly contesting and challenging automated decisions. However, it still places a substantial impetus on the data subject to protect its own rights, and only indirectly challenges the knowledge logic of computational empiricism. Explainability may even be harmful if entrenching automated decision making and narrowing the types of reasons to be given for decisions and thus the grounds for contestation. Being subjected to automated decisions without understanding how or why that decision was made may be problematic. But receiving inadequate automated explanations without recourse might be worse. In that format, 'explanation' similarly becomes a legal vector for proliferating decision automation. However, other approaches appear to more directly challenge data science's epistemological dominance.

Hildebrandt, for instance, outlines several concepts that challenge data science's program of quantification – what she describes as the 'overcomplete datafication of anything and everything based on the idea that the mathematics that grounds all these machines reveals the ultimate layer of a hidden reality.'[110] She offers a right to human non-computability built on the philosophical principle of indeterminate identity. This is not necessarily a return to older opacity paradigms of privacy, a relocation of the black box to the level of the individual, but rather a mechanism for limiting certain classes of knowledge claims. It is achieved in her vision through 'agonistic' machine learning systems, capable of 'demanding that companies or governments that base decisions on machine learning must explore and enable alternative ways of datafying and modelling the same event, person or action.'[111] Agonistic systems would demonstrate how each act of computation relies on a specific system of measurements, representations, and analytics. Rather than challenging human computability wholesale, the project highlights how humans can be computed in multiple ways, in order to 'ward off monopolistic claims about the 'true' or the 'real' representation of human beings'.[112] This is arguably also a form of explanation and accountability, but one targeted more appropriately at the decision making process itself rather than the relationship between input and outcome (i.e. counter-factual analysis). In other words, an explanation mechanism informing how to transcend the solution space of any particular decision. This type of mechanism is appealing because it more directly addresses the construction of the human subject as information pattern. It also recognises the significance of representations in a manner similar to technical approaches like 'disentangling the factors of variation' in order to produce 'disentangled representations'.[113] What this manner of legal intervention would actually look like requires more legal, conceptual, and technical thinking however. But it at least offers a pathway to envisioning new projects built for a technical age that sees humans in a very different way than our existing legal systems insist on seeing them.

---

[109] Andrew Selbst and Julia Powles, 'Meaningful Information and the Right to Explanation' (2017) 7(4) *International Data Privacy Law* 233; S Wachter, B Mittelstadt, and L Floridi, 'Why a Right to Explanation of Automated Decision-Making Does Not Exist in the General Data Protection Regulation' (2017) 7 *International Data Protection Law* 76; B Goodman and S Flaxman, 'European Union Regulations on Algorithmic Decision-Making and a "Right to Explanation"' (2017) 38 *AI Magazine* 50.
[110] Mireille Hildebrandt, 'Privacy as Protection of the Incomputable Self: From Agnostic to Agonistic Machine Learning' <https://papers.ssrn.com/sol3/papers.cfm?abstract_id=3081776>.
[111] Ibid.

**Conclusion**

This article has attempted to outline some emerging challenges produced by a profound new technical capacity. Lawmakers now have to contend with a radically extrapolated enlightenment philosophy insisting that measuring everything and drawing knowledge from those measurements is the path to truth. Permitting such measurement and classification without limitation risks generating a fair, transparent, and non-prejudicial totalitarianism. This has been described as the 'insanity proper to logic', whereby measuring everything, logic conceives 'a world in which all things are relative, makes itself absolute, and denying the whole of nature, establishes its own artifices.'[114] Addressing this reality means clarifying how computer vision systems and data science applications are merely apparatuses, combining and computing symbols that encode the world a particular way, according to particular programs. Without intervention at that level, digital systems risk simultaneously constituting both the world and the dominant understanding of it.[115] As Flusser reminds us, if humans cease to decode technical images and instead project them unencoded back onto the world 'out there', the world 'itself becomes like an image – a context of scenes, of states of things.'[116] Critiquing data science therefore means exposing the difference between world and the 'world state', and challenging the idea that such systems access a 'hidden reality' instead of producing and operationalise a *para*-reality built from *para*-visual representations.

---

[112] Ibid.
[113] Valentin Thomas et al, 'Independently Controllable Factors' (2017) *arXiv:1708.01289*.
[114] Jacques-Allain Miller, 'Jeremy Bentham's Panoptic Machine' (1987) 41 *October* 3, 7.

Computer vision has been presented as one vector of that knowledge logic, and one target of a new kind of legal thinking. However, this is certainly not the case for computer vision alone. It is relevant for any data science application translating the real world into the symbolic register of computation. The goal is to open up space for legal ideas that introduce contingency and automated un-decidability into those translations that are building the environments ordering social experience. Such projects seem more and more imperative, especially in the face of a new form of computational empiricism designed for 'looking at people'.

---

[115] See e.g. Leif Weatherby, 'Digital Metaphysics: The Cybernetic Idealism of Warren McCulloch' (2018) 20(1) *The Hedgehog Review* (online) < https://iasc-culture.org/THR/THR_article_2018_Spring_Weatherby.php>
[116] Flusser, above n 19, 10.